\documentclass[12pt]{article}
\usepackage{times}
\usepackage{latexsym}                      
\usepackage[dvips]{graphics}

\def\ls#1{\dimen0=\fontdimen6\the\font
          \lineskip=#1\dimen0 \advance\lineskip.5\fontdimen5\the\font
          \advance\lineskip-\dimen0 \lineskiplimit=.9\lineskip
           \baselineskip=\lineskip  \advance\baselineskip\dimen0}


\date{April 10, 2006}

\title{Rainfall Advection using  Velocimetry by Multiresolution Viscous Alignment \thanks{This material  is  supported  in part  by  NSF ITR 0121182 and DDDAS 0540259.}}

\author{Sai Ravela\\
Earth, Atmospheric and Planetary Sciences \\
 Virat Chatdarong\\ 
Civil and Environmental Engineering\\
Massachusetts Institute of Technology\\
{\sf ravela@mit.edu}}

\begin{document}
\maketitle
\pagestyle{headings}
\ls{2}

\begin{abstract}

An algorithm  to estimate motion from satellite  imagery is presented.
Dense displacement  fields are computed from  time-separated images of
of significant convective activity using a Bayesian formulation of the
motion estimation problem.   Ordinarily this motion estimation problem
is ill-posed; there are far too many degrees of freedom than necessary
to  represent  the motion.   Therefore,  some  form of  regularization
becomes  necessary and  by imposing  smoothness and  non-divergence as
desirable  properties  of  the  estimated displacement  vector  field,
excellent  solutions are  obtained.   Our approach  provides a  marked
improvement over  other methods in  conventional use.  In  contrast to
correlation based approaches, the  displacement fields produced by our
method are dense, spatial consistency of the displacement vector field
is  implicit, and  higher-order  and small-scale  deformations can  be
easily  handled.   In  contrast  with optic-flow  algorithms,  we  can
produce solutions  at large separations of  mesoscale features between
large time-steps or where the deformation is rapidly evolving. 

\end{abstract}
\section{Introduction}

 Environmental  data  assimilation is  the  methodology for  combining
imperfect  model  predictions  with  uncertain  data  in  a  way  that
acknowledges their respective uncertainties.  The proper framework for
state        estimation       includes       sequential~\cite{gelb74},
ensemble-based~\cite{evensen03}                                     and
variational~\cite{lorenc86,courtier97}   methods.   

 The  difficulties   created  by  improperly   represented  error  are
particularly  apparent in mesoscale  meteorological phenomena  such as
thunderstorms, squall-lines, hurricanes, precipitation, and fronts. We
are  particularly  interested  in  rainfall  data-assimilation,  where
rainfall  measurements from  satellite  data, radar  data, or  in-situ
measurements are used to  condition a rainfall model. Such conditional
simulations are  valuable both for producing estimates  at the current
time (nowcasting), as well as for short-term forecasting.

There  are a  countless number  of  models developed  to simulate  the
rainfall process. In  general, there are two types  of models that can
deal with spatial and temporal characteristics of rainfall.  The first
category is the meteorological model or the quantitative precipitation
forecasting model.   It involves a large, complex  set of differential
equations seeking to represent complete physical processes controlling
rainfall and other weather related variables. Examples of these models
include     the     fifth-generation     Mesoscale     Model     (MM5)
\cite{chen01a,chen01b,grell93}, the step-mountain Eta coordinate model
\cite{black94,black93,rogers96}, and the Regional Atmospheric Modeling
System (RAMS) \cite{orlandi04,pielke92}, etc.   The second type is the
spatiotemporal  stochastic rainfall  model. It  aims to  summarize the
spatial and  temporal characteristics  of rainfall by  a small  set of
parameters             \cite{copertwait91,khaliq96,moradkhani05,onof00,
over96,rodriguez88}.  This  type of model usually  simulates the birth
and decay of  rain-cells and evolve them through  space and time using
simple physical  descriptions.  Despite significant  differences among
these rainfall  models, the concept of  propagating rainfall through
space and time are relatively similar.

The  major  ingredient  required  to  advect rainfall  is  a  velocity
field. Large spatial-scale (synoptic) winds are inappropriate for this
purpose for  a variety of reasons.   Ironically, synoptic observations
can be  sparse to  be used directly  and although  synoptic-scale wind
analyses  produced from  them (and  models) do  produce  dense spatial
estimates,  such estimates  often do  not contain  variability  at the
meso-scales of interest.  The  motion of mesoscale convective activity
is  a natural source  for velocimetry.   Indeed, there  exist products
that deduce  ``winds'' by estimating the motion  of temperature, vapor
and other fields evolving in time~\cite{velden97,velden05}.

In this paper,  we present an algorithm for  velocimetry from observed
motion from satellite observations such  as GOES, AMSU, TRMM, or radar
data  such  as  NOWRAD.    This  algorithm  follows  from  a  Bayesian
formulation  of   the  motion   estimation  problem,  where   a  dense
displacement  field   is  estimated  from  two   images  of  cloud-top
temperature of  rain-cells separated  in time.  Ordinarily,  the motion
estimation problem is ill-posed, because the displacement field has far
too many degrees of freedom  than the motion.  Therefore, some form of
regularization  becomes  necessary  and  by  imposing  smoothness  and
non-divergence  as desirable properties  of the  estimated displacement
vector field solutions can be obtained.

This  approach  provides  marked  improvement over  other  methods  in
conventional use.   In contrast  to correlation based  approaches used
for deriving  velocity from GOES imagery, the  displacement fields are
dense, quality  control is implicit, and  higher-order and small-scale
deformations  can  be  easily  handled. In  contrast  with  optic-flow
algorithms~\cite{nagel83,heeger88}, we  can produce solutions at  large separations of
mesoscale features  between large time-steps or  where the deformation
is rapidly evolving.

After  formulating  the  motion  estimation problem  and  providing  a
solution, we extend the algorithm using a multi-resolution procedure.
The  primary advantage  of a  multi-resolution approach  is  to produce
displacement fields  quickly. The secondary advantage  is to structure
the estimation  homotopically; coarse or  low-frequency information is
used  first  to  produce  velocity estimates  over  which  deformation
adjustments from finer-scale structures is superposed. The result is a
powerful  algorithm for  velocimetry  by alignment.  As  such, it  is
useful  in  a  variety  of  situations  including,  for  example,  (a)
estimating winds,  (b) estimating  transport of tracers,  (c) Particle
Image Velocimetry, (d) Advecting Rainfall models etc.

\section{Related Work}
\label{rw}

There are two dominant  approaches to computing flow from observations
directly. The  first is correlation-based  and the second is  based on
optic flow. 

In correlation based  approaches~\cite{lawton83}, a region of interest
(or patch)  is identified in the  first image and  correlated within a
search window in the second image.   The location of the best match is
then used  to compute a displacement  vector. When the  input image or
field  is tiled,  possibly overlapping,  and regions  of  interest are
extracted  from  each tile  location,  the  result  is velocimetry  at
regular  intervals  and  is  most  commonly used  for  Particle  Image
Velocimetry  (PIV).   In certain  instances  it  is  useful to  define
interest-points or salient features around which to extract regions of
interest. In particular,  if the field has many  areas with negligible
spatial variability, then matches  are undefined. As a quality control
measure then, matching is restricted only to those regions of interest
that have interesting variability, or interest points.

There     are    several     disadvantages     to    correlation-based
approaches. First, by  construction it is assumed that  the entire ROI
purely translates from one image to  the other. This is not always the
case, but  is a reasonable  approximation when the right  length scale
can  be found.   However, when  higher-order deformations  (shears for
example) are present, correlation  based approaches cannot be expected
to  work well.  Second,  correlation based  approaches  assume that  a
unique match can  be found in a way that  is substantially better than
correlation  elsewhere.  This  is   only  true  if  the  features  are
well-defined and identified.  Third,  there is no implicit consistency
across  regions of  interest in  correlation-based  flow.  Neighboring
regions of  interest can  and often do  match at wildly  different and
inconsistent locations. This calls for a significant overhead in terms
of quality control. Fourth, it is not clear how the search window size
(that is  the area over which a  region of interest is  matched in the
subsequent frame) is determined. This window size varies both in space
(as the velocity  varies spatially) and time (as  velocity varies with
time). A  larger search window  portends a larger probability  to miss
the  real  target, and  a  smaller search  window  can  lead to  false
negatives or false positives.  Finally, where interest points are used
as a preprocessing step to correlation, the velocity field produced is
necessarily sparse, and therefore,  leaves hanging the question of how
to produce dense flow fields. Our proposed algorithm handles all these
issues in a simple and direct way.

More   closely   related   to   the   proposed   approach   is   optic
flow~\cite{nagel83,heeger88}. This method arises from what is known as
the  brightness   constraint  equation,   which  is  a   statement  of
conservation  of   brightness  (intensity)  mass,   expressed  by  the
continuity equation evaluated at each pixel or grid node of $X$.
\begin{equation}
\label{cont}
\frac{\partial X}{\partial t}+ {\bf q}\cdot \nabla X  = 0
\end{equation}

Here $X$ is  the brightness or intensity scalar field  and ${\bf q}$ a
displacement vector-field. Solutions to the optic flow equation can be
formulated using the well-known method by~\cite{nagel83}, which can be
stated as a solution to the following system of equations:

\begin{equation}
\label{optic}
 (\nabla X) (\nabla X)^T {\bf q}   = -(\nabla X)\frac{\partial X}{\partial t}
\end{equation}

The right-hand side is completely determined from a pair of images and
the  coefficient or  stiffness matrix  on  the left-hand  side is  the
second-derivative of  the auto correlation  matrix, also known  as the
windowed second-moment  matrix, or Harris interest  operator, which is
sensitive to ``corners'' in an image. This formulation arises directly from a quadratic formulation, which can in turn be synthesized from a Bayesian formulation under a Gaussian assumption. Thus, we can write that we seek to minimize 
\begin{equation}
J({\bf q})  =  || X({\bf r - q}) - Y||
\end{equation}

Then solve this problem via the Euler-Lagrange equation:
\begin{eqnarray}
\label{el}
\frac{\partial J({\bf q})}{\partial {\bf q}} & = &  \nabla X|_{{\bf r-q}} (X({\bf r - q}) - Y) = 0
\end{eqnarray}

The solution is obtained by {\em linearizing} (\ref{el}), that is,
\begin{eqnarray}
 \nabla X|_{{\bf r-q}} (X({\bf r}) - \nabla X \cdot {\bf q} - Y) & =& 0\nonumber \\
\label{optfinal} \nabla X(\nabla X)^T {\bf q}& = & - \nabla X( Y - (X({\bf r}))
\end{eqnarray}
There are  several disadvantages to this algorithm.   First, much like
correlation   with  feature   detection,   equation~\ref{optfinal}  is
evaluated at pixels where the second-moment matrix is full-rank, which
corresponds  to locations  where features  are present.   There  is no
clear way  of propagating information obtained at  sparse locations to
locations where direct computation of displacement is not possible due
to poor conditioning of the second-moment matrix. For the same reason,
it cannot handle tangential  flows. The brightness constraint equation
can only represent flows along brightness streamlines. When tangential
motion  is present,  detected motion  at extreme  ends a  moving curve
cannot be  propagated easily into  the interior.  Our  method provides
some  degree  of  spatial   smoothness  common  in  geophysical  fluid
transport,  and  uses  regularization  constraints to  propagate  flow
information to nodes where feature strengths are weak.

Second, the linearization implicit in (\ref{optfinal}) precludes large
displacements;  structures must be  closely overlapping  in successive
images,  which   can  also  be  seen  from   the  continuity  equation
(\ref{cont}).   Therefore,  this method  is  very  useful for  densely
sampled motion, such as  ego-motion resulting from a moving, jittering
camera, but  is not as useful  for sparsely sampled  flow arising from
structures moving in  a scene.  In the latter  case, to ameliorate the
effects  of large  expected displacement,  multi-resolution approaches
have been  proposed.  Even so, much  like determining the  size of the
search window in correlation, determining the number of resolutions is
an ad-hoc procedure.  Our method can handle large displacements and we
also propose  a multi-resolution approach, but  the primary motivation
there is improved computational speed.

\section{Velocimetry by Field Alignment}

The main approach consists of solving a nonlinear quadratic estimation
problem for  a field of  displacements. Solutions to this  problem are
obtained  by  regularizing  the  an  ill-posed  inverse  problem.  The
material presented  in this section  is derived directly from  work by
Ravela~\cite{ravela06b}, and Ravela  et al.~\cite{ravela06a}.  Here we
reformulate  their   original  formulation  to   allow  only  position
adjustments.

\begin{figure}[htbp!]
\centering 
\scalebox{1}{\includegraphics{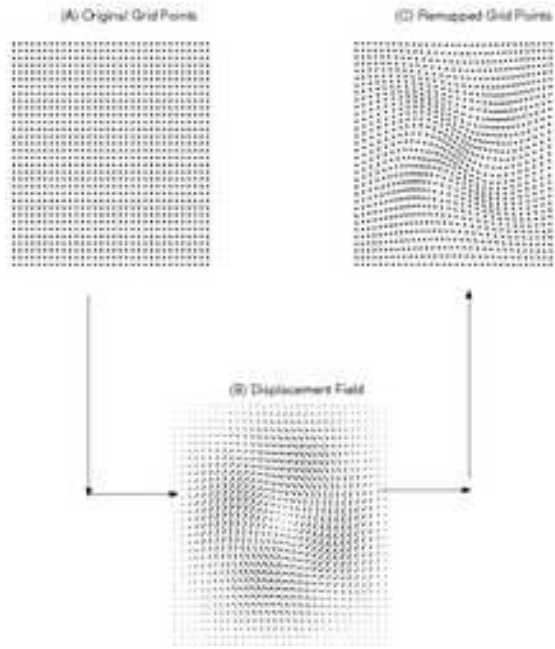}}
\caption{A graphical illustration of  field alignment. State vector on
a discretized  grid is  moved by  deforming its grid  (${\bf r}$)  by a
displacement (${\bf q}$).}
\label{dispfig}
\end{figure}

 To make this  framework more explicit it is  useful to introduce some
notation.    Let  $   X=X({\bf   r})  =   \{X[\underline{r}^T_1]\ldots
X[\underline{r}^T_m]\}$  be  the first  image,  written  as a  vector,
defined over a spatially  discretized computational grid $\Omega$, and
${\bf r^T}  = \{\underline{r}_i =  (x_i,y_i)^T, i\in \Omega\}$  be the
position indices.  Let  ${\bf q}$ be a {\em  vector} of displacements,
that is  ${\bf q^T} =  \{\underline{q}_i = (\Delta  x_i,\Delta y_i)^T,
i\in \Omega\} $. Then the  notation $ X({\bf r}-{\bf q})$ represents {
\em displacement} of  $X$ by ${\bf q}$.  The  displacement field ${\bf
q}$  is real-valued, so  $ X({\bf  r}-{\bf q})$  must be  evaluated by
interpolation if  necessary. It is  important to understand  that this
displacement field represents a  warping of the underlying grid, whose
effect   is   to   move   structures   in  the   image   around,   see
Figure~\ref{dispfig}.

In a probabilistic  sense, we may suppose that  finding ${\bf q}$ that
has the  maximum a posteriori probability in  the distribution $P({\bf
q}|{\cal X},{\cal  Y})$ is  appropriate.  Without loss  of generality,
${\cal X}$ is a random variable corresponding to the image or field at
a given time and ${\cal Y}$ is  random variable for a field at a future
time.  Using  Bayes rule we obtain $P(Q = {\bf  q}|{\cal X}=X,{\cal Y}=Y) \propto P({\cal Y} = Y,{\cal X} = X|{\bf
q})P({\bf  q})$. If  we make  a Gaussian  assumption of  the component
densities, we can write:

\begin{equation}
    P(X,Y|{\bf       q})      =
\frac{1}{(2\pi)^{\frac{n}{2}}\left|R\right|^{\frac{1}{2}}}e^{-\frac{1}{2}\left(Y
- X\left({\bf r}-{\bf  q}\right)\right)^TR^{-1}\left(Y - X\left({\bf r}-{\bf
q}\right)\right)}  
\end{equation}
This  equation says  that the  observations separated  in time  can be
related using a Gaussian model  to the displaced state X({\bf r}- {\bf
q}), where X({\bf  r}) is defined on the original  grid, and ${\bf q}$
is a  displacement field.  We  use the linear observation  model here,
and therefore,  $Y = HX({\bf  r - q})+  \eta, \eta \sim  N(0,R).$.  We
should  emphasize here  that the  observation vector  is  fixed.  It's
elements are always defined from  the original grid. In fully observed
fields,  H  is  an  identity  matrix, and  for  many  applications  R,
reflecting the noise in the field,  can also be modeled as an identity
matrix.

\begin{equation}
\label{dprior}   P({\bf   q})    =    \frac{1}{C}e^{-L({\bf   q})}
\end{equation}

This  equation specifies a  {\em displacement  prior}.  This  prior is
constructed  from  an energy  function  $L({\bf  q})$ which  expresses
constraints  on  the  displacement  field.  The  proposed  method  for
constructing $L$ is drawn from the nature of the expected displacement
field.  Displacements can be represented as smooth flow fields in many
fluid  flows  and smoothness   naturally  leads  to  a  Tikhonov  type
formulation  ~\cite{tikhonov77} and,  in particular,  $L({\bf  q})$ is
designed  as  a  gradient   and  a  divergence  penalty  term.   These
constraints, expressed in quadratic form are:

\begin{equation}
\label{const}
L({\bf q}) = \frac{w_1}{2}\sum\limits_{j\in\Omega} {\bf tr}\{[\nabla { \underline{q}_{j}}][\nabla { \underline{q}_{j}}]^T\} + \frac{w_2}{2}\sum\limits_{j\in\Omega} [\nabla\cdot { \underline{q}_{j}}]^2
\end{equation} 

In Equation~\ref{const},  ${\bf q}_j$ refers  to the $j^{th}$ grid  index and
{\bf  tr}   is  the  trace.   Equation~\ref{const}  is   a  {\em  weak
constraint},   weighted  by  the   corresponding  weights   $w_1$  and
$w_2$.  Note  that  the constant  C  can  be   defined  to  make
Equation~\ref{dprior}  a proper  probability  density. In  particular,
define $Z({\bf q}) = e^{-L({\bf q})}$  and define $C = \int\limits_{{\bf q}}
Z({\bf q})d{\bf q}$. This integral exists and converges.

With  these definitions  of probabilities,  we  are in  a position  to
 construct an objective by evaluating the log probability.  We propose
 a solution using Euler-Lagrange equations.  Defining ${\bf p = r - q}$These can be written as:
\begin{eqnarray}
\label{disp}
\frac{\partial J}{\partial {\bf q}} &=&  \nabla X|_{{\bf p}}H^TR^{-1}\left(H\; X\left({\bf p}\right) - Y\right) + \frac{\partial L}{\partial {\bf q}}=0
\end{eqnarray}

Using the regularization constraints  (~\ref{disp}) at a node
$i$ now becomes:
\begin{equation}
\label{dispfinal}
w_1 \nabla^2  \underline{q}_{i} + w_2 \nabla(\nabla\cdot { \underline{q}_{i}}) + \left[\nabla X^f{^T}|_{{\bf p}}H^TR^{-1}\left(H\left[X^f\left({\bf p}\right)\right] - Y\right)\right]_i =0
\end{equation}

Equation~\ref{dispfinal}  is  the  field  alignment  formulation.   It
introduces  a forcing  based on  the residual  between the  model- and
observation-fields.  The  constraints on the  displacement field allow
the    forcing    to    propagate    to   a    consistent    solution.
Equation~\ref{dispfinal}   is   also   non-linear,   and   is   solved
iteratively, as a Poisson equation.  During each iteration ${\bf q}$
is computed  by holding  the forcing term  constant.  The  estimate of
displacement at  each iteration is then  used to deform a  copy of the
original  forecast model-field  using bi-cubic  interpolation  for the
next iteration.   The process is  repeated until a  small displacement
residual is  obtained, the misfit with observations  does not improve,
or  an iteration  limit  is  reached.  Upon  convergence,  we have  an
aligned  image  $X({\bf  \hat{p}})$, and a displacement field ${\bf \hat{q}} = \sum\limits_{d=1}^N q^{(d)}$, for individual displacements $q^{(d)}$ at iterations $d=1\ldots D$

\subsection{Multi-resolution Alignment and Velocimetry}

The convergence of solution  to the alignment equation is super-linearly
dependent   on    the   expected   displacement    between   the   two
fields. Therefore,  it is  desirable to solve  it in  a coarse-to-fine
manner,  which serves  two principal  advantages.  The  first,  as the
following  construction will  show, is  to substantially  speed-up the
time   to   alignment   because   decimated   (or   coarse-resolution)
representations of a pair  of fields has smaller expected displacement
than a  pair at finer resolution. 

Second, decimation  or resolution reduction
also implies  that finer structure or higher  spatial frequencies will
be  attenuated.  This  smoothness in  the  coarsened-field intensities
directly translates to  smoothness in flow-fields using (~\ref{disp}).
Thus,  a coarse-to-fine  method  for alignment  can incrementally  add
velocity   contributions   from   higher-frequencies,   that   is   it
incrementally    incorporates   higher-order   variability    in   the
displacement  field.   Many of  the  advantages  of a  multi-resolution
approach have been previously explored in the context of visual motion
estimation,  including the famous  pyramid algorithm  and architecture
for matching and flow and our implementation borrows from this central
idea.

\begin{figure}[htbp!]
\centering \scalebox{1}{\includegraphics{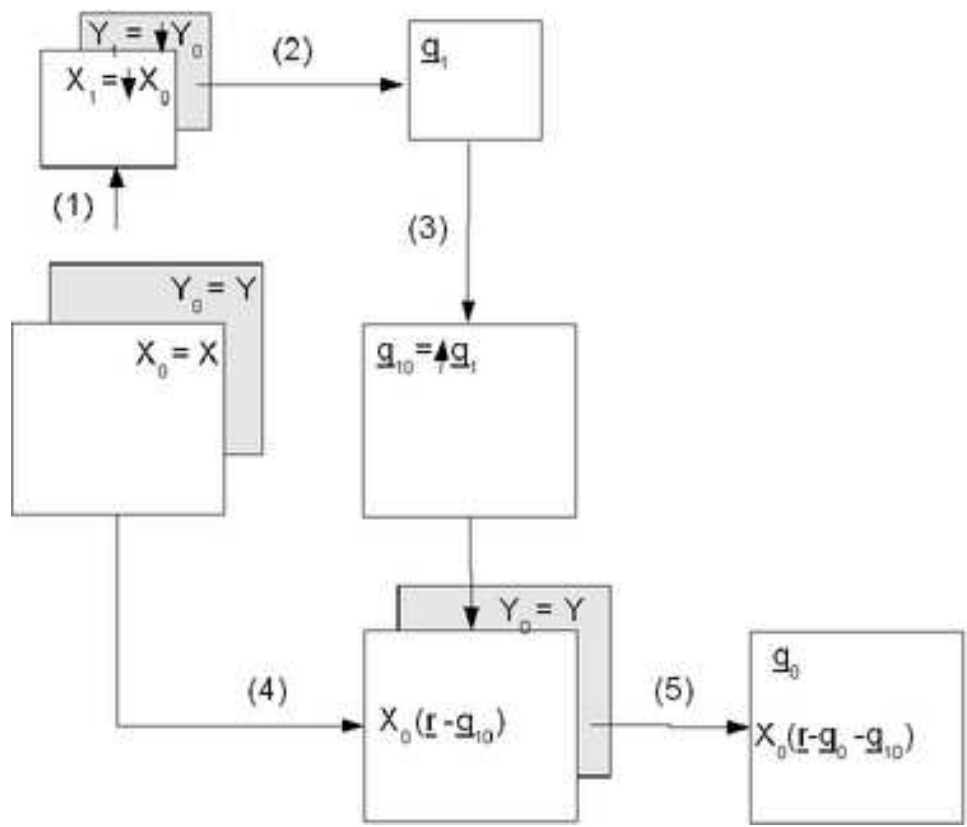}}
\caption{The multi-resolution algorithm is shown for two-levels and requires five steps, labeled (1) through (5). See text for explanation.}
\label{multires}
\end{figure}

The  multi-resolution algorithm  is depicted  in Figure~\ref{multires}
for  two  levels.  The  input  images $X$  and  $Y$  are decimated  to
generate coarse  resolution images $X_1$ and  $Y_1$ respectively (step
1). Let us  suppose that this scaling is by a  factor of $0<s<1$ (most
commonly $s=0.5$). Displacement is  computed for this level first, and
let us call this ${\bf  \hat{q}_1}$ (step 2).  This displacement field
is   downscaled  by   a  factor   of  $s$,   using   simple  (bicubic)
interpolation, to produce a prior estimate of displacement at level 0,
written ${\bf \hat{q}_{10}} = {s^{-1}\bf \hat{q}_{0}}(s^{-1}{\bf r}) $
(step 3). The source image at level  0, that is $X_0 = X$ is displaced
by ${\bf \hat{q}_{10}}$ (step 4)  and thus $X({\bf r - \hat{q}_{10}})$
is  aligned  with  $Y_0$  to  produce a  displacement  estimate  ${\bf
\hat{q}_{0}}$ (step 5).  The  total displacement relating source image
$X$   with  target   field  $Y$   is  simply   ${\bf   \hat{q}_{0}  +
\hat{q}_{10}}$.  Multiple levels of resolution can be implemented from
this framework recursively.

\section{Example}

\begin{figure}[ht!]
\centering \scalebox{0.8}{\includegraphics{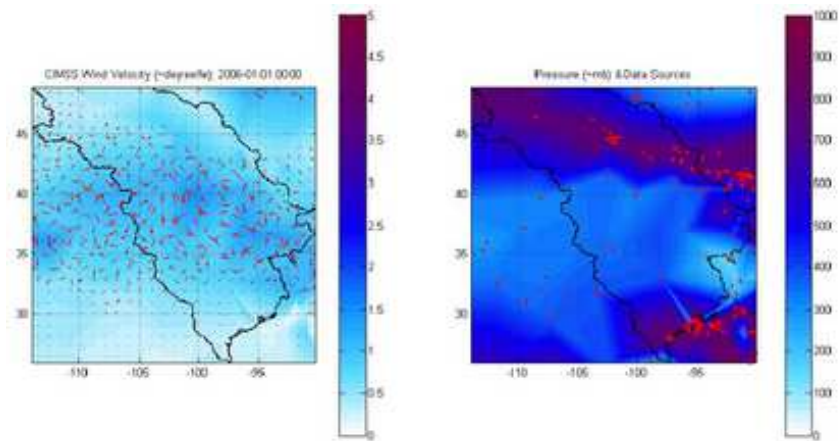}}
\caption{CIMSS Winds derived from GOES data at 2006-04-06-06Z (left) and pressure (right). The velocity vectors are sparse and contain significant divergence.}
\label{cimss1}
\end{figure}

\begin{figure}[ht!]
\centering\scalebox{0.8}{\includegraphics{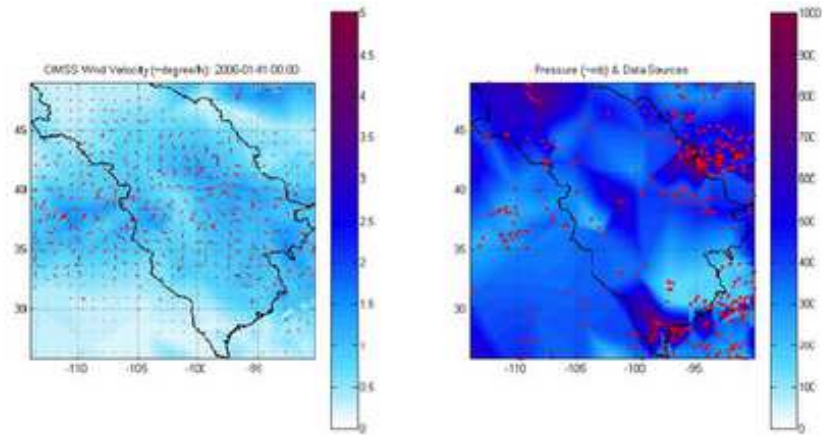}}
\caption{CIMSS Winds derived from GOES data at 2006-04-06-09Z (left) and pressure (right). The velocity vectors are sparse and contain significant divergence.}
\label{cimss2}
\end{figure}

The  performance of  this algorithm  is illustrated  in  a velocimetry
computation.    To   compare,  we   use   CIMSS  wind-data   satellite
data~\cite{velden05},     depicted     in    Figure~\ref{cimss1},
 and Figure~\ref{cimss2} obtained from
CIMSS analysis on 2006-06-04 at  06Z and 09Z respectively.  CIMSS
wind-data is  shown over the US  great plains, and  were obtained from
the 'sounder.'   The red  dots indicate the  original location  of the
data.  The  left subplot shows  wind speed (in degree/hr).   The right
ones show pressure, and the location of raw measurements in red.

It     can      be     seen      in     the     maps      shown     in
Figure~\ref{cimss1} and Figure~\ref{cimss2}   that
current method to produce  winds generate sparse vectors and, further,
has  substantial  divergence.   Whilst  this  can be  thought  of  as
accurately representing turbulence, in reality these vectors are more
likely  the result of  weak quality  control. The  primary methodology
used  here is to  identify features  in an  image, extract  regions of
interest around  them and search  for them in subsequent  frames. This,   by  definition   produces  sparse   velocity  estimates
(features  are  sparse),  leaving  unanswered  how  to  systematically
incorporate  appropriate  spatial   interpolation  functions  for  the
velocity. Since  regions of interest are essentially  treated as being
statistically  independent,  mismatches  can  produce  widely  varying
displacement   vectors.   Such   mis-matches  can   easily   occur  in
correlation based approaches when  the features are not distinguishing
or substantial deformations occur from one time to another in a region
of   interest.    A  more   detailed   discussion   is  presented   in
Section~\ref{rw}.

\begin{figure}[ht!]
\centering \scalebox{1}{\includegraphics{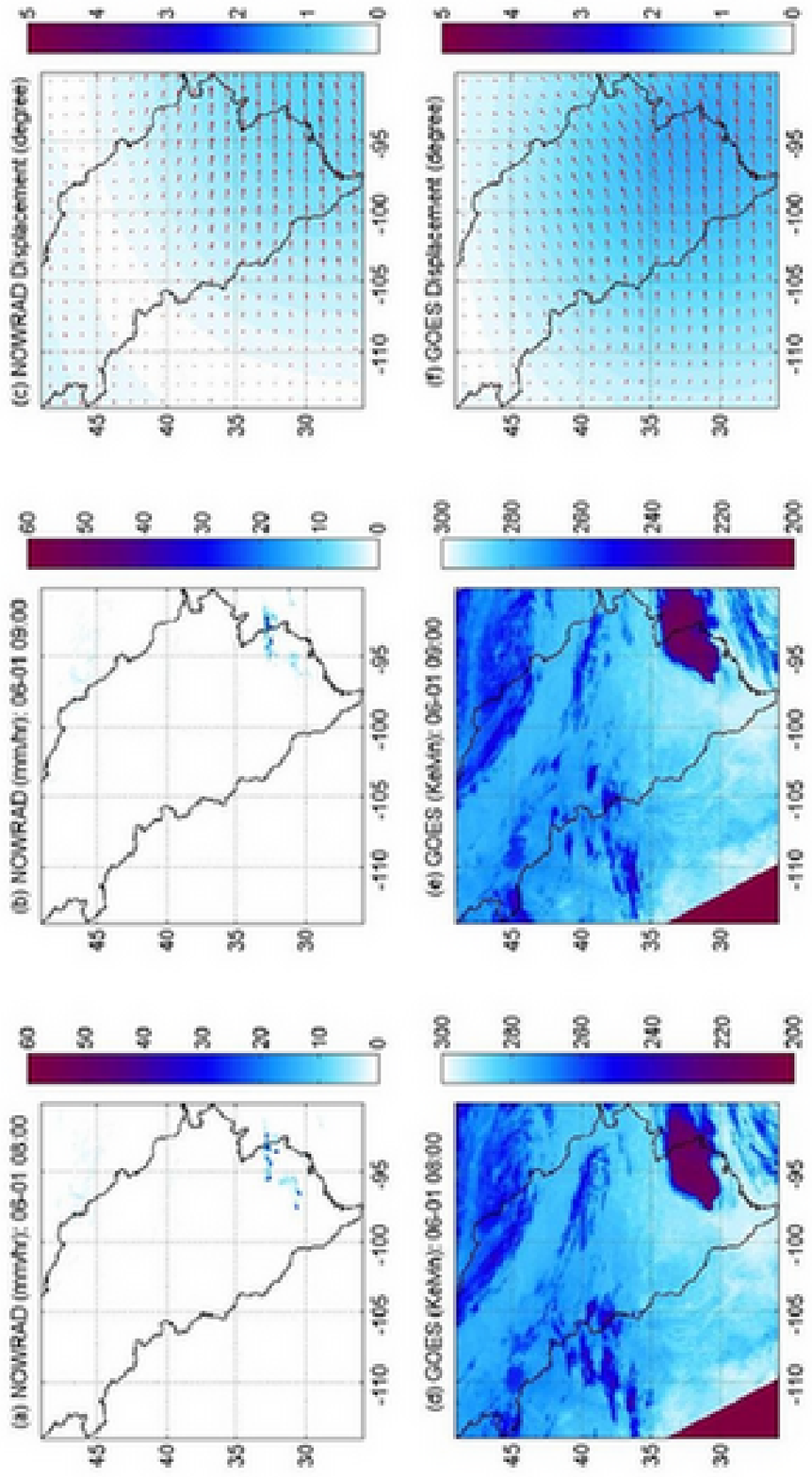}}
\caption{Deriving velocimetry information from satellite observations,
Nexrad (top), GOES (bottom). See text for more information.}
\label{goes}
\end{figure}

In  contrast,  our method  produces  dense  flow  fields, and  quality
control   is  implicit   from   regularization  constraints.    Figure
~\ref{goes}(a,b) shows a pair of NOWRAD images at 2006-06-01-0800Z and
2006-06-01-0900Z  respectively,   and  the  computed   flow  field  in
Figure~\ref{goes}(c).   Similarly,  Figure~\ref{goes}(d,e,f) show  the
GOES images and velocity from  the same time frame over the deep convective rainfall region in the Great Plains example. The velocities are
in  good  agreement with  CIMSS  derived  winds  where magnitudes  are
concerned, but  the flow-fields are smooth and  visual confirmation of
the alignment provides convincing evidence that they are correct.

\section{Conclusions}

  Our method is a Bayesian  perspective of the velocimetry problem. It
has several distinct advantages: (a) It  is useful for a wide range of
observation modalities.  (b) Our approach does not require features to
be identified for computing  velocity. This is a significant advantage
because  features  cannot often  be  clearly  delineated,  and are  by
definition sparse.   (c) Our approach implicitly  uses quality control
in  terms  of smoothness,  and  produces  dense  flow-fields. (d)  our
approach   can   be  integrated   easily   with  current   operational
implementations, thereby making this effort more likely to have a real
impact.   Finally,   it  should  be  noted   that  the  regularization
constraint in  field alignment  is a weak  constraint and  the weights
determine how strongly the  constraints influence the flow field.  The
constraint in $L$ is modeled as  such because we expect the fluid flow
to be smooth.  From a regularization point of view, there can be other
choices~\cite{wabha80} as well. The proposed  method can be used for a
variety  of  velocimetry  applications  including PIV,  velocity  from
tracer-transport, and velocity from GOES and other satellite data, and
an application of this is  to advect rain-cells produced by a rainfall
model, with realistic wind-forcing.

\bibliographystyle{plain}
\bibliography{phase}
\end{document}